\documentclass[copyright,creativecommons]{eptcs}
\providecommand{\event}{GRAPHITE 2012} % Name of the event you are submitting to
\usepackage{breakurl}             % Not needed if you use pdflatex only.
\usepackage{prelude}
\usepackage{amsmath}

\newcommand{\Title}{Efficient Instantiation of Parameterised Boolean Equation Systems to Parity Games}
\newcommand{\ShortTitle}{\Title}
\newcommand{\Author}{
\begin{tabular}{c@{\qquad\qquad}c}
Gijs Kant\thanks{Gijs Kant is sponsored by the NWO under grant number 612.000.937 (VOCHS).}
& Jaco van de Pol\\
\texttt{\footnotesize kant@cs.utwente.nl}  & \texttt{\footnotesize vdpol@cs.utwente.nl}
\end{tabular}
\medskip
\\{\footnotesize Formal Methods \& Tools}
\\{\footnotesize University of Twente}
\\{\footnotesize Enschede, The Netherlands}
}
\newcommand{\AuthorRunning}{G. Kant \& J.C. van de Pol}

\begin{document}
%\mainmatter
\title{\Title}
\author{\Author}
\def\titlerunning{\ShortTitle}
\def\authorrunning{\AuthorRunning}
\maketitle

\begin{abstract}
Parameterised Boolean Equation Systems (PBESs) are sequences of Boolean fixed point equations with data variables,
used for, e.g., verification of modal \MUCALC formulae for process algebraic specifications with data.

Solving a PBES is usually done by instantiation to a Parity Game and then solving the game.
Practical game solvers exist, but the instantiation step is the bottleneck.

We enhance the instantiation in two steps.
First, we transform the PBES to a Parameterised Parity Game (PPG), a PBES with each equation either conjunctive or disjunctive.
Then we use \LTSMIN, that offers transition caching, efficient storage of states and both distributed and symbolic state space generation, for generating the game graph.
To that end we define a language module for \LTSMIN, consisting of an encoding of variables with parameters into state vectors, a grouped transition relation and a dependency matrix to indicate the dependencies between parts of the state vector and transition groups.

Benchmarks on some large case studies, show that the method speeds up the instantiation significantly and decreases memory usage drastically.
\end{abstract}


\section{Introduction}
Parameterised Boolean Equation Systems (PBESs) are sequences of fixed point equations with data variables.
They form a very expressive formalism for encoding a wide range of problems, such as 
the verification of modal \MUCALC formulae \cite{kozen1983:results, bradfield2001:modal} for process algebraic specifications with data (see, e.g., \cite{groote2005:modelchecking, groote2005:parameterised}) and
checking for (branching) bisimilarity of process equations \cite{chen2007:equivalence}.

PBESs have been described extensively in \cite{groote2005:parameterised}.
A method for solving PBESs directly has been presented \cite{groote2005:modelchecking}, 
but usually PBESs are solved by first instantiating the system to a plain Boolean Equation System (BES)
and then solving the BES.
Instantiation of PBESs is described in \cite{vandam2008:instantiation, ploeger2011:verification}, where clever rewriters and enumeration of quantifier expressions play an important role. 
We focus on instantiation to a Parity Game (PG), which is a restricted BES with equations that are either conjunctive or disjunctive. 
Although no polynomial time algorithm for solving parity games is known (however, the problem is known to be in $\text{NP}\cap\text{co-NP}$), effective parity game solvers exist (see, e.g., \cite{friedmann2009:solving}), especially when the alternation depth is low, and the instantiation step is currently the bottleneck of the whole procedure for many practical cases.

There are clear similarities between instantiation of PBESs and state space generation, a well known problem in model checking. In both, an abstract description gives rise to a large graph, which requires efficient storage of the generated graph. Also, in both we often have that the description consists of a combination of reasonably independent components or equations. This `locality' can be used to speed up the generation of successor nodes. 
Inspired by these similarities, we apply in this paper optimisations from model checking to the PBES instantiation problem, devising a more efficient method.
We use \LTSMIN, a language independent toolset for state space exploration which enables efficient state space generation and offers both symbolic exploration tools based on Binary Decision Diagrams (BDDs) and distributed exploration tools (see, e.g., \cite{blom2010:ltsmin}). The tools make use of knowledge about the dependencies for better efficiency, which can be specified for every language in a separate language module.
Instantiating PBESs to parity games in our enhanced method has two phases: 
\begin{enumerate}\noitemsep
\item Transforming the PBES into an equivalent system that consists of expressions that are either purely conjunctive or purely disjunctive. We call such a system a \emph{Parameterised Parity Game} (PPG).
The result of this operation is that any instantiation of the PPG will result directly in a parity game. 
\item Instantiating the PPG to a PG using \LTSMIN. To this end this we defined a PBES language module for \LTSMIN, 
in which we specify a state vector representation of instantiated PBES variables (and the corresponding node in the generated game graph) and the dependencies between (parts of) the equations and the parts of the state vector. 
\end{enumerate}

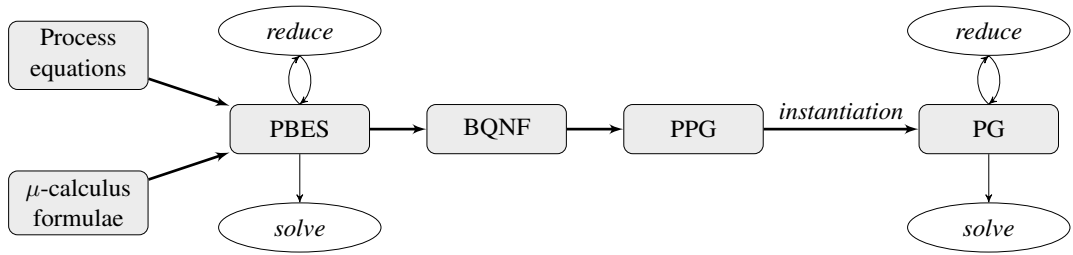
\begin{figure}[tbh]
\centering
\scalebox{.85}{
  \tikzstyle{textblock} = [rectangle, text width=5em, text centered, minimum height=2em]
\tikzstyle{block} = [rectangle, draw, fill=gray!15, text width=5em, text centered, rounded corners, minimum height=2em]
\tikzstyle{process} = [ellipse, draw, text width=4em, text centered, minimum height=2em]
\tikzstyle{arrow} = [draw, -stealth']
\tikzstyle{transformation} = [draw, -latex', very thick]

\begin{tikzpicture}[node distance=8em, auto, bend angle=50]
  \node [block] (mcrl) {Process equations};
  \node [below of=mcrl, node distance=3em] (dummy) {};
  \node [block, below of=dummy, node distance=3em] (mcalc) {$\mu$-calculus formulae};

  \node [block, right of=dummy, node distance=9em] (pbes) {PBES};
  \node [block, right of=pbes] (bqnf) {BQNF};
  \node [block, right of=bqnf] (ppg) {PPG};
  \node [block, right of=ppg, node distance=12em] (bes) {PG};
  \node [process, below of=pbes, node distance=4em] (solvepbes) {\itshape solve};
  \node [process, above of=pbes, node distance=4em] (reducepbes) {\itshape reduce};
  \node [process, below of=bes, node distance=4em] (solvebes) {\itshape solve};
  \node [process, above of=bes, node distance=4em] (reducebes) {\itshape reduce};

  \path [transformation] (mcrl) -- (pbes);
  \path [transformation] (mcalc) -- (pbes);
  \path [transformation] (pbes) -- (bqnf);
  \path [transformation] (bqnf) -- (ppg);
  \path [transformation] (ppg) edge node {\itshape instantiation} (bes);
  \path (pbes.north) edge [arrow, bend left] (reducepbes.south);
  \path (reducepbes.south) edge [arrow, bend left] (pbes.north);
  \path (bes.north) edge [arrow, bend left] (reducebes.south);
  \path (reducebes.south) edge [arrow, bend left] (bes.north);  
  \path [arrow] (pbes) -- (solvepbes);
  \path [arrow] (bes) -- (solvebes);
\end{tikzpicture}
}
\caption{Overview of the verification approach, consisting of various transformations, an instantiation step, and 
available reductions and solvers.}
\label{fig:verification}
\end{figure}

An overview of the method is shown in Figure~\ref{fig:verification}.
We consider PBESs in \emph{Bounded Quantifier Normal Form} (BQNF), which is a subset of all PBESs, but any PBES can be rewritten automatically to a system in BQNF with the same solution.
PBESs and their normal forms are described in Section~\ref{section:background}.
The contributions of this article are the transformation from BQNF to PPG and the instantiation from PPG to PG.
Both steps are not trivial. We will explain here where the obstacles lie.

In general, each system of PBES equations in BQNF can be transformed automatically into a system consisting of equations in PPG while preserving the solution. An equation can be transformed to PPG by introducing fresh equations for subexpressions and replacing the subexpressions by the corresponding variable.
However, it is important not to separate quantifiers from the expressions that restrict the data elements that have to be considered, so called \emph{bounds}. If a bound for a quantifier over an infinite data sort is replaced by a variable, the instantiator might generate an infinite number of successors for a node in the game graph.
See Section~\ref{section:transformation} for our solution.

For the instantiation step we implemented a PBES language module for \LTSMIN using the Partitioned Interface for the Next State function (\PINS). This includes partitioning each PBES equation into \emph{transition groups} and defining a \emph{dependency matrix} that specifies the dependencies between transition group and parts of the state vector.
We then have a high-performance instantiation tool that offers both distributed and symbolic generation of a parity game.
This requires some delicacy, as splitting a formula too much may result in infinite computation (as in the transformation phase) and not splitting enough could result in a dependency matrix that is too dense, which ruins the effect of transition caching and symbolic computation.
The implementation is described in Section~\ref{section:instantiation}.

In Section~\ref{section:experiments} we present performance results for a number of case studies, comparing our sequential, distributed and symbolic implementations based on \LTSMIN to the existing PBES instantiation tools in the \MCRLTWO toolset. In almost all cases memory usage is orders of magnitude better for our tool. In all cases also the execution time is much better.

\section{Background}\label{section:background}

In this section we will treat PBESs, normal forms for PBESs, and Parity Games.

\subsection{PBES}

\begin{definition}
\emph{Predicate formulae} $\varphi$ are defined by the following grammar:
\[ \varphi \Coloneqq b \mid \propvar{X}(\vec{e}) \mid \neg \varphi \mid \varphi \oplus \varphi \mid \mathsf{Q} d \oftype D \suchthat \varphi \]
where $\oplus \in \set{\land, \lor, \impl}$, $\mathsf{Q} \in \set{\forall, \exists}$, 
$b$ is a data term of sort $\Bool$, 
$\propvar{X} \in \X$ is a predicate variable, 
$d$ is a data variable of sort $D$, and 
$\vec{e}$ is a vector of data terms.
We will call any predicate formula without predicate variables a \emph{simple formula}.
We denote the class of predicate formulae $\PF$.
\end{definition}

\begin{definition}
A \emph{First-Order Boolean Equation} is an equation of the form:
\[ \sigma \propvar{X}(\vec{d} \oftype D) = \varphi \]
where $\sigma \in \set{\mu, \nu}$ is a minimum ($\mu$) or maximum ($\nu$) fixed point operator,
$\vec{d}$ is a vector of data variables of sort $D$, and
$\varphi$ is a predicate formula.
\end{definition}

\begin{definition}
A \emph{Parameterised Boolean Equation System (PBES)} is a sequence of First-Order Boolean Equations:
\[ \eqsys = (\sigma_1 \propvar{X}_1(\vec{d_1} \oftype D_1) = \varphi_1) 
   \eqsep \dotsc 
   \eqsep (\sigma_n \propvar{X}_n(\vec{d_n} \oftype D_n) = \varphi_n) \]
\end{definition}

The semantics and solution of PBESs are described in, e.g., \cite{groote2005:parameterised}.
We say that two equation systems $\eqsys_1$ and $\eqsys_2$ are equivalent, written as $\eqsys_1 \equiv \eqsys_2$, if they have the same solution for every variable that occurs in both systems.

We adopt the standard limitations: expressions are in positive form (negation occurs only in data expressions) and every predicate variable occurs exactly once as the left hand side of an equation.
A PBES that contains no quantifiers and parameters is called a \emph{Boolean Equation System} (BES).
A finitary PBES can be \emph{instantiated} to a BES by expanding the quantifiers to finite conjunctions
or disjunctions and substituting concrete values for the data parameters. Every instantiated PBES variable $\propvar{X}(\vec{e})$ should then be read as a BES variable ``$\propvar{X}(\vec{e})$''.

A one-to-one mapping can be made from a BES to an equivalent \emph{parity game} if the BES has only expressions that are either conjunctive or disjunctive. The parity game is then represented by a game graph with nodes that represent variables with concrete parameters and edges that represent dependencies.
Parity games will be further explained in Section~\ref{section:paritygames}.
To make instantiation of a PBES to a parity game more directly 
we will preprocess the PBES to a format that only allows expressions to be either conjunctive or disjunctive. 
This format is a normal form for PBESs that we call
the \emph{Parameterised Parity Game}, defined as follows:

\begin{definition}
A PBES is a \emph{Parameterised Parity Game} (PPG) if every right hand side of an equation is a formula of the form:
\begin{align*} 
          \Land_{i \in I} f_i \land
          \Land_{j \in J} \forall_{\vec{v} \in D_j} \suchthat \big( g_j \impl \propvar{X}_j(\vec{e_j}) \big) \quad
\vert \quad
          \Lor_{i \in I} f_i \lor
          \Lor_{j \in J} \exists_{\vec{v} \in D_j} \suchthat \big( g_j \land \propvar{X}_j(\vec{e_j}) \big). 
\end{align*}
where $f_i$ and $g_j$ are simple boolean formulae, and $\vec{e_j}$ is a data expression. $I$ and $J$ are finite (possibly empty) index sets.
\end{definition}

The expressions range over two index sets $I$ and $J$. The left part is a conjunction (or disjunction) of simple expressions $f_i$ that can be seen as conditions that should hold in the current state. The right part is a conjunction (or disjunction) of a quantified vector of variables for next states $\propvar{X}_j$ with parameters $\vec{e_j}$, guarded by simple expression $g_j$.

Before transforming arbitrary PBESs to PPGs we first define another normal form on PBESs to make the transformation easier.
This normal form can have an arbitrary sequence of bounded quantifiers as outermost operators and has
a conjunctive normal form at the inner.
We call this the Bounded Quantifier Normal Form (BQNF):

\begin{definition}
A First-Order Boolean formula is in \emph{Bounded Quantifier Normal Form (BQNF)} if it has the form:
\begin{align*}
\mathsf{BQNF} \Coloneqq &\quad
    \forall {\vec{d} \in D} \suchthat b \impl \mathsf{BQNF}
\quad \vert \quad
    \exists {\vec{d} \in D} \suchthat b \land \mathsf{BQNF}
\quad \vert \quad
    \mathsf{CONJ} \\
\mathsf{CONJ} \Coloneqq &\quad
    \Land_{k \in K} f_k \land
    \Land_{i \in I} \forall_{\vec{v} \in D_I} \suchthat \big( g_i \impl \mathsf{DISJ}^i \big) \\
\mathsf{DISJ}^i \Coloneqq &\quad
    \Lor_{\ell \in L_i} f_{i\ell} \lor
    \Lor_{j \in J_i} \exists_{\vec{w} \in D_{ij}} \suchthat \big( g_{ij} \land \propvar{X}_{ij}(\vec{e_{ij}}) \big)
\end{align*}
where $b$, $f_k$, $f_{i\ell}$, $g_i$, and $g_{ij}$ are simple boolean formulae, and $\vec{e_{ij}}$ is a data expression. 
$K$, $I$, $L_i$, and $J_i$ are finite (possibly empty) index sets.
\end{definition}

This BQNF is similar to \emph{Predicate Formula Normal Form} (PFNF), defined elsewhere\footnote{
A transformation to PFNF is implemented in the \texttt{pbesrewr} tool and documented at
\url{http://www.win.tue.nl/mcrl2/wiki/index.php/Parameterised_Boolean_Equation_Systems}.},
in that quantification is outermost and in that the core is a conjunctive normal form. 
However, unlike PFNF, BQNF allows bounds on the quantified variables (hence bounded quantifiers), and universal quantification
is allowed within the conjunctive part and existential quantification is allowed within the disjunctive parts.
These bounds are needed to avoid problems when transforming to PPG. 
Consider the expression 
$(\forall i \oftype \Nat \suchthat (i < 5) \impl \propvar{Y}(i) ) 
\lor ( \exists j \oftype \Nat \suchthat (j < 3) \land \propvar{Z}(j) )$.
Rewriting to PFNF (moving the quantifiers outward) results in 
$\exists j \oftype \Nat \suchthat \forall i \oftype \Nat \suchthat 
((i < 5) \impl Y(i)) 
\lor ((j < 3) \impl Z(j))$. 
Rewriting that expression to PPG would split the expression such that the initial expression is
$\exists j: \Nat \suchthat \propvar{X_1}(j)$ ($\propvar{X_1}$ is a newly introduced variable for the equation with the remainder of the expression as right hand side), which would result in an infinite disjunction when instantiating the PPG. 
BQNF allows the original expression to be rewritten to 
$\exists j \oftype \Nat \suchthat (j < 3) \land \forall i \oftype \Nat \suchthat (i < 5) \impl (\propvar{Y}(i) \lor \propvar{Z}(j))$ with the bounds close to the quantifiers, which allows to split the expression after the bound, preventing the instantiation to result in an infinite expression.
Requiring that a system is specified in BQNF does not limit the expressiveness, as each PBES can be transformed into a equivalent system in PFNF that has the same solution and PFNF is a subset of BQNF. 

The translation from process algebraic specifications in \MCRLTWO and \MUCALC formulae to PBESs is given in \cite{groote2005:modelchecking} and is illustrated by the following example.
Throughout the paper we expect the reader to know process algebras and to be able to read \MCRLTWO specifications\footnote{See \url{http://mcrl2.org} for documentation on the \MCRLTWO language.}.

\begin{example}[Buffer]\label{example:buffer}
Consider the specification of a simple buffer with a capacity of 2.
\begin{quote}
\begin{tabular}{l@{}l@{\hspace{3pt}}l}
$\sort\ $ & \multicolumn{2}{@{}l@{}}{$D = \struct\ d_1 \mid d_2;$ \quad $\act\ \action{r_1}, \action{s_4} \oftype D;$} \\
$\proc\ $ & $\procname{Buffer}(q \oftype \container{List}(D)) = $
          & $\displaystyle\sum_{d \oftype D} \ (\#q < 2) \guards \action{r_1}(d) \suchthat \procname{Buffer}(q \append d)$ \\
        & & $\ \ \, + \ (q \neq []) \guards \action{s_4}(\head(q)) \suchthat \procname{Buffer}(\tail(q));$ \\
$\init\ $ & $\procname{Buffer}([]);$
\end{tabular}
\end{quote}
The specification consists of sort and action definitions, process specifications where alternatives are in summands and an initial state.
On the first line an enumerated data sort $D$ is introduced with data values $d_1$ and $d_2$, and the actions $\action{r_1}$ and $\action{s_4}$ are specified, both having a data parameter of type $D$.
A process $\procname{Buffer}$ is specified that has a data parameter $q$, which is a list of elements of type $D$. The process consists of \emph{summands}, separated by the $+$-operator. Each summand may start with a summation over a data set, followed by a guard that is closed with a $\guards$, then an action, followed by a call to the process that describes the behaviour after the action, typically a recursive call to the process itself with different parameters.

The first summand specifies that any element $d$ can be added to $q$ by the action $\action{r_1}(d)$ 
if the size of the internal buffer $q$ is smaller than $2$. 
The second summand specifies that if $q$ is not empty,
elements can be popped by the action $\action{s_4}(\head(q))$.
The initial state of the system is the $\procname{Buffer}$ process with an empty list in this case, which models that initially the buffer is empty.

We can check the specification for absence of deadlock, which is expressed in \MUCALC as follows:
\begin{center}
$\always{\true^\ast}\possibly{\true}\true \quad (\text{which is syntactic sugar for: }\; \nu X \suchthat \possibly{\true}\true \land \always{\true}X )$
\end{center}
which reads: after any sequence of actions ($\always{\true^\ast}$), always some action is enabled ($\possibly{\true}\true$).
Satisfaction of the formula by the specification, translated to a PBES, looks as follows:
\begin{quote}
\begin{tabular}{l@{}l@{\hspace{3pt}}l}
$\sort\ $ & \multicolumn{2}{@{}l@{}}{$D = \struct\ d_1 \mid d_2;$} \\
$\pbes\ $ & $\nu \propvar{X}(q \oftype \container{List}(D)) = $
          & $(q \neq []) \lor (\#q < 2)$ \\
        & & $\land \ (q \neq []) \impl \propvar{X}(\tail(q))$ \\
        & & $\land \ \displaystyle\forall_{d \in D} \suchthat (\#q < 2) \impl \propvar{X}(q \append d);$ \\
$\init\ $ & $\propvar{X}([]);$
\end{tabular}
\end{quote}\medskip
This PBES is true if from the initial state $\propvar{X}([])$ an element can be added to $q$ if
$\#q$ is smaller than $2$, an element can be popped from $q$ if it is not empty and 
any of these actions is enabled ($q \neq []$ or $\#q < 2$, which is obviously true for any $q$).
The same has to hold
for the successor states ($\propvar{X}$ with an element added to, respectively popped from $q$ as parameter).
The solution of the PBES is $\ttrue$.
\end{example}

\begin{rem*}
The equation system in the example above is already a PPG, which is no coincidence as any system when combined with the absence of deadlock property will result in a PBES in PPG form because of the form of the formula: a conjunction of ``we can do an action now'' (a disjunctive expression without recursion) and ``for all possible actions the property holds in all next states'' (universal quantification with recursion).
Note that checking the absence of deadlock property is almost the same as standard reachability analysis.
\end{rem*}

\begin{definition}[Block]\label{def:block}
A PBES is divided into \emph{blocks}, which are subsequences of equations with the same fixed point operator
such that subsequent equations with the same fixed point operator belong to the same block.
\end{definition}

%\begin{example}
%For a PBES \;
%$  \nu \propvar{X_1} = \varphi_1 \eqsep 
%    \nu \propvar{X_2} = \varphi_2 \eqsep
%    \mu \propvar{X_3} = \varphi_3 \eqsep 
%    \nu \propvar{X_4} = \varphi_4 ,
%$ \;
%equations \propvar{X_1} and \propvar{X_2} form block 1, equation \propvar{X_3} forms block 2, and equation \propvar{X_4} forms block 3.
%\end{example}

\subsection{Parity Games}\label{section:paritygames}

A \emph{parity game} 
is a game between two players, player \Eloise (also called Eloise or player \emph{even}) and player \Abelard (also called Abelard or player \emph{odd}), where
each player owns a set of places. On one place a token is placed that can be moved 
by the owner of the place to an adjacent place.
The parity game is represented as a graph. We borrow notation from \cite{bradfield2001:modal} and \cite{mazala2002:infinite}.

\begin{definition}[Parity Game]
A \emph{parity game} is a graph $\G = \tuple{V, E, \VEloise, \VAbelard, v_I, \Omega}$, with
\begin{itemize}\noitemsep
\item $V$ the set of vertices (or places or states);
\item $E \oftype V \times V$ the set of transitions;
\item $\VEloise \subseteq V$ the set of places owned by player \Eloise;
\item $\VAbelard \subseteq V$ the set of places owned by player \Abelard;
\item $v_I \in V$ the initial state of the game; 
\item $\Omega \oftype V \to \Nat$ assigns a priority $\Omega(v)$ to each vertex $v \in V$;
\end{itemize}
where $\VEloise \cup \VAbelard = V$ and $\VEloise \cap \VAbelard = \emptyset$.
\end{definition}

The nodes in the graph represent the places and correspond to the instantiated variables from the equation system. 
The edges represent possible moves of the token (initially placed on $v_I$) and encode dependencies between variables.
A node does not necessarily have outgoing transitions, i.e., deadlock nodes are allowed.
In the parity game, player \Eloise owns the nodes that represent disjunctions, player \Abelard the nodes that represent conjunctions.

The node priorities correspond to the number of the block
to which the corresponding variable belongs (see Def.~\ref{def:block}), such that
variables in earlier blocks have lower priorities, $\nu$-blocks have even priorities, $\mu$-blocks have odd priorities and the earliest $\mu$-block has priority $1$.
%Table~\ref{table:pg-bes} 
The following table shows an intuitive overview of the relations between BESs and parity games.
%\begin{table}[tb]
%\caption{Overview of the correspondence between (P)BESs and parity games.}
%\label{table:pg-bes}
\begin{center}
\begin{tabular}{c@{\quad}c@{\qquad}l}
\toprule
&$\nu$ blocks   & Even priorities (0, 2, 4, \ldots) \\
&$\mu$ blocks   & Odd priorities (1, 3, 5, \ldots) \\
&$\lor$, $\exists$, $\possibly{}$        & Player \Eloise, {$\exists$}loise, Even, Prover \\
&$\land$, $\forall$, $\always{}$         & Player \Abelard, {$\forall$}belard, Odd, Refuter \\
\bottomrule
\end{tabular}
\end{center}
%\end{table}
\medskip
The values \ttrue ($\true$) and \tfalse ($\false$) are represented as a node with priority 0, player \Abelard and a transition to itself, and a node with priority 1, player \Eloise and a transition to itself, respectively.

A \emph{play} in the game is a finite path $\pi = v_0 v_1 \cdots v_r \in V^+$ ending
in a deadlock state $v_r$
or an infinite path $\pi = v_0 v_1 \cdots \in V^\omega$ such that  
$(v_i, v_{i+1}) \in E$ for every $v_i \in \pi$.
Priority function $\Omega$ extends to plays in the following way:
$\Omega(\pi) = \Omega(v_0) \Omega(v_1) \cdots$.
$\Inf(\rho)$ returns the set of values that occur infinitely often in a sequence $\rho$.

\begin{definition}[Winner of a play]
Player \Eloise is the winner of a play $\pi$ if %
\begin{itemize}\noitemsep
\item $\pi$ is a finite play $v_0 v_1 \cdots v_r \in V^+$ and
$v_r \in \VAbelard$ and no move is possible from $v_r$; or
\item $\pi$ is an infinite play and $\min(\Inf(\Omega(\pi)))$,
the minimum of the priorities that occur infinitely often
in $\pi$,
is even.
This is called the \emph{min-parity condition}.
\end{itemize}%
\end{definition}
\begin{definition}[Strategy]
A (memoryless) \emph{strategy} for player $a$ is a function $f_a \oftype V_a \to V$.
A play $\pi = v_0 v_1 \cdots$ is \emph{conform} to $f_a$ if for every $v_i \in \pi$, \;
$v_i \in V_a \impl v_{i+1} = f_a(v_i)$.
\end{definition}
\begin{definition}[Winner of the game]
Player \Eloise is the \emph{winner} of the game if and only if there exists a winning strategy 
for player \Eloise, i.e., from the initial state every play conforming to the strategy will be
won by player \Eloise.
\end{definition}

%The value $\ttrue$ may be encoded as a node with priority $0$ and a self loop, corresponding to the equation
%$\nu \propvar{X} = \propvar{X}$ in the first $\nu$-block. The value $\tfalse$ may be encoded as a node with priority $1$
%and a self loop, corresponding to the equation $\mu \propvar{Y} = \propvar{Y}$ in the first $\mu$-block.

The model checking problem is encoded as a PBES (see \cite{groote2005:modelchecking}) which is instantiated to a parity game (see \cite{ploeger2011:verification}) such that 
player \Eloise is the winner of the game iff the property holds for the system.

\vspace*{-\medskipamount}
\paragraph{Solving Parity Games}
Solving a parity game means finding a winning strategy for one of the players. 
Various algorithms exist, such as the recursive algorithm by Zielonka \cite{zielonka1998:infinite} and Small Progress Measures by Jurdzi\'{n}ski \cite{jurdzinski2000:spm}, with a multi-core implementation in \cite{vdpol2008:multicore}. An overview and performance comparison of the algorithms are given in \cite{friedmann2009:solving}.

\section{Transformation from BQNF to Parameterised Parity Games}\label{section:transformation}

In order to automatically transform a PBES to a PPG, we define a transformation function $s$ from BQNF to PPG.
The transformation rewrites expressions that contain both conjunctions and disjunctions to equivalent expressions that are either conjunctive or disjunctive, by introducing new equations for certain subformulae and substituting calls to the new equations for these subformulae in the original expression.
The function $t$ below replaces an expression by a call to a new equation if the expression is not already a variable instantiation. The function $t'$ introduces a new equation for an expression if needed.
\begin{align*}
t(\propvar{X},\vec{d},\varphi) &\eqdef
\begin{cases}
\varphi                              
  & \text{ if } \varphi \text{ is of the form } \propvar{X}'(\vec{e}), \\
\propvar{X}(\vec{d})
  & \text{ otherwise; }
\end{cases}\\
t'(\sigma,\propvar{X},\vec{d},\varphi) &\eqdef
\begin{cases}
\emptyseq
  & \text{ if } \varphi \text{ is of the form } \propvar{X}'(\vec{e}), \\
s( \sigma \propvar{X}(\vec{d}) = \varphi )  
  & \text{ otherwise. }
\end{cases}
\end{align*}%
For brevity, we leave out the types of the parameters.
A tilde is used to introduce a fresh variable: $\fresh{\propvar{X}}$.
For equation system
$ \eqsys = (\sigma \propvar{X}_1(\vec{d}_1) = \xi_1) \eqsep \ldots \eqsep (\sigma \propvar{X}_n(\vec{d}_n) = \xi_n), $
with each $\xi_i$ in BQNF, the translation to PPG is defined as follows:
\begin{center}
\scalebox{1}{
\begin{tabular}{l@{\hspace{2pt}}l}
$s\big(\eqsys \big)$ & $\eqdef
    s\big( \sigma \propvar{X}_1(\vec{d}_1) = \xi_1 \big) \eqsep \ldots \eqsep s\big( \sigma \propvar{X}_n(\vec{d}_n) = \xi_n \big)$
    \defsep
    
$s\big(\sigma \propvar{X}(\vec{d}) = f \big)$ & $\eqdef
    \sigma \propvar{X}(\vec{d}) = f$
    \defsep
$s\big(\sigma \propvar{X}(\vec{d}) = \forall \vec{v} \suchthat b \impl \varphi \big)$ & $\eqdef
    \Big( \sigma \propvar{X}(\vec{d}) = \forall \vec{v} \suchthat b \impl t(\fresh{\propvar{X}},\vec{d}+\vec{v},\varphi) \Big)$
    \\&
    $\qquad
    t'(\sigma,\fresh{\propvar{X}},\vec{d}+\vec{v},\varphi)$
    \defsep
$s\big(\sigma \propvar{X}(\vec{d}) = \exists \vec{v} \suchthat b \land \varphi \big)$ & $\eqdef
    \Big( \sigma \propvar{X}(\vec{d}) = \exists \vec{v} \suchthat b \land t(\fresh{\propvar{X}},\vec{d}+\vec{v},\varphi) \Big)$
    \\&
    $\qquad
    t'(\sigma,\fresh{\propvar{X}},\vec{d}+\vec{v},\varphi)$
    \defsep
    
$s\big(\sigma \propvar{X}(\vec{d}) = 
      \Land_{k \in K} f_k$ \\
      \qquad\qquad $\land \Land_{i \in I} (\forall_{\vec{v}_i} \suchthat g_i \impl \varphi_i) \big)$ & $\eqdef
    \Big( \sigma \propvar{X}(\vec{d}) = 
      \Land_{k \in K} f_k$ \\
    & \qquad\qquad\quad $\land
      \Land_{i \in I} \big(\forall_{\vec{v}_i} \suchthat g_i \impl t(\fresh{\propvar{X}}_i,\vec{d}+\vec{v}_i,\varphi_i)\big) \Big)$
    \\
    &
    $\qquad
    t'(\sigma,\fresh{\propvar{X}}_1,\vec{d}+\vec{v}_1,\varphi_1)
    \eqsep
    \ldots
    \eqsep
    t'(\sigma,\fresh{\propvar{X}}_m,\vec{d}+\vec{v}_m,\varphi_m)$
    \defsep
    
$s\big(\sigma \propvar{X}(\vec{d}) = 
      \Lor_{k \in K} f_k$ \\
      \qquad\qquad $\lor \Lor_{i \in I} (\exists_{\vec{v}_i} \suchthat g_i \land \varphi_i) \big)$ & $\eqdef
    \Big( \sigma \propvar{X}(\vec{d}) = 
      \Lor_{k \in K} f_k$ \\
    & \qquad\qquad\quad $\lor
      \Lor_{i \in I} \big(\exists_{\vec{v}_i} \suchthat g_i \land t(\fresh{\propvar{X}}_i,\vec{d}+\vec{v}_i,\varphi_i)\big) \Big)$
    \\&
    $\qquad
    t'(\sigma,\fresh{\propvar{X}}_1,\vec{d}+\vec{v}_1,\varphi_1)
    \eqsep
    \ldots
    \eqsep
    t'(\sigma,\fresh{\propvar{X}}_m,\vec{d}+\vec{v}_m,\varphi_m)$
\end{tabular}
}
\end{center}
with 
$I = 1 \ldots m$, $\vec{v} \cap \vec{d} = \emptyset$ (variables in $\vec{v}$ do not occur in $\vec{d}$), $b$, $f$, $f_k$, $g_i$ are simple formulae, $\varphi$, $\varphi_i$ are formulae that may contain predicate variables.

\begin{prop}
The transformation $s$ is solution preserving, i.e., for any $\eqsys$ in BQNF, $s(\eqsys) \equiv \eqsys$: 
bound variables $\propvar{X}(d)$ 
have the same solution in $s(\eqsys)$ as in $\eqsys$.
\begin{proof}
Every change made by $s$ to an equation $\sigma \propvar{X} = \xi$ is a substitution of a 
subexpression $\varphi$ by a fresh variable $\fresh{\propvar{X}}$, 
while adding at the same time a new equation $\sigma \fresh{\propvar{X}} = \varphi$ in the same block as $\propvar{X}$.
We can apply \emph{backward substitution} (using \cite[Lemma 18]{groote2005:parameterised})
$s(\sigma \propvar{X} = \xi)[\fresh{\propvar{X}} \becomes \varphi]$ for every substitution caused by the transformation
to get the original equation system (plus an unused equation $s(\sigma \fresh{\propvar{X}} = \varphi)$ for every
fresh variable \fresh{\propvar{X}}).
From that we can conclude that $s(\eqsys) \equiv \eqsys$.
\end{proof}
\end{prop}

\begin{example}[Example of the transformation]
We combine the buffer from Example~\ref{example:buffer} with the property that in every state both $\action{r_1}$ and $\action{s_4}$ actions
are enabled:
\[ \nu \propvar{X} \suchthat (\exists_{d \oftype D} \suchthat \possibly{\action{r_1}(d)}\propvar{X}) \land
   (\exists_{d \oftype D} \suchthat \possibly{\action{s_4}(d)}\propvar{X}) \]
The resulting PBES has an equation which does not conform to the PPG form, but is in BQNF:
\begin{quote}
\begin{tabular}{l@{}l@{\hspace{3pt}}l@{\hspace{3pt}}l}
$\sort\ $ & \multicolumn{3}{@{}l@{}}{$D = \struct\ d_1 \mid d_2;$} \\
$\pbes\ $ & $\nu \propvar{X}(q \oftype \container{List}(D)) =$ &
          $\big( \exists_{d \oftype D} \suchthat (\#q < 2) \land \propvar{X}(q \append d) \big)$\\
        & &
          $\land \big( \exists_{d \oftype D} \suchthat (\head(q) = d) \land (q \neq []) \land \propvar{X}(\tail(q)) \big);$\\
$\init\ $ & $\propvar{X}([]);$
\end{tabular}
\end{quote}\medskip
The transformation $s$ replaces both conjuncts by a fresh variable and adds equations for these variables with 
the substituted expression as right hand side, resulting in equations:
\begin{quote}
\begin{tabular}{l@{}l@{\hspace{3pt}}l@{\hspace{3pt}}l@{\hspace{3pt}}l}
%$\sort$ & \multicolumn{3}{@{}l@{}}{$D = \struct d_1 \mid d_2; \quad \eqn N = 2;$} \\
$\pbes\ $ & $\nu \propvar{X}(q \oftype \container{List}(D))$ & $=$ 
        & $\propvar{X_1}(q) \land \propvar{X_2}(q);$ \\
        & $\nu \propvar{X_1}(q \oftype \container{List}(D))$ & $=$ 
        & $\exists_{d \oftype D} \suchthat (\#q < 2) \land \propvar{X}(q \append d);$ \\
        & $\nu \propvar{X_2}(q \oftype \container{List}(D))$ & $=$ 
        & $\exists_{d \oftype D} \suchthat (\head(q) = d) \land (q \neq []) \land \propvar{X}(\tail(q));$ \\

%$\init$ & $\propvar{X}([]);$
\end{tabular}
\end{quote}
The first equation is purely conjunctive, while that latter two equations are (guarded) disjunctive.
\end{example}

\section{Instantiation of Parameterised Parity Games}\label{section:instantiation}

We view the instantiation of PPGs to Parity Games as generating a transition system, where states are predicate variables with concrete parameters and transitions are dependencies, specified by the right hand side of the corresponding equation in the PPG.

\begin{example}
Consider the equation:
\[ \nu \propvar{X}(d \oftype D) = (d > 0 \land d < 10) \impl \propvar{X}(d-1) \land \propvar{X}(d+1) \]
If $\propvar{X}(5)$ is the initial value, its successors are $\propvar{X}(4)$ and $\propvar{X}(6)$,
so the graph starts with a node owned by player \Abelard representing $\propvar{X}(5)$ with transitions to nodes $\propvar{X}(4)$ and $\propvar{X}(6)$.
\end{example}

\subsection{\LTSMIN}

We use the tool \LTSMIN to generate a parity game given a PPG. \LTSMIN is a language independent tool for state-space generation \cite{blom2010:ltsmin}. Different language-modules are available, which are connected to different exploration algorithms through the so-called \PINS-interface. This interface allows for certain language-independent optimisations, such as transition caching and distributed generation (see \cite{blom2009:bridging}), and an efficient compressed storage of states in a tree database (see \cite{blom2009:database}). Also symbolic reachability analysis is possible, where the state space is stored as a Binary Decision Diagram (BDD) \cite{blom2008:symbolic}.

\subsubsection{Partitioned Interface for the Next State function}

\LTSMIN uses a Partitioned Interface for the Next State function (\PINS), where
states are represented as a vector $\tuple{x_1, x_2, \dotsc, x_M}$ with size $M$ that is fixed for the whole system (to be determined statically).
These values are stored in a globally accessible table, so that the states can also be represented as a vector of integer indices $\tuple{i_1, i_2, \dotsc, i_M}$.
The \PINS interface functions on this level of integer vectors, so that each tool can really be language-independent.
Throughout the text we will often use value vectors instead of index vectors for better readability.

For a system with a state vector of $M$ parts, the universe of states is $S = \Nat^M$.
For each language module a \emph{transition function} $\nextst \oftype S \to \powerset{S}$ has to be defined that computes the set of successor states for a given state.
This transition relation is preferrably split into \emph{transition groups} 
in order to reflect the compositional structure of the system,
by defining a function
$\groupnext \oftype S \times \Nat \to \powerset{S}$ that computes successors for state $s$ as defined in group $k$.
Suppose we have $K$ transition groups. $\nextst$ can then be defined as 
\[ \nextst(s) = \bigcup_{k=1}^{K}\groupnext(s,k)
\]

\subsubsection{Dependence}
An important optimisation comes from the observation that not all parts of the state vectors
are relevant in every transition group.
To indicate the relevant parts of the vector for each of the transition groups, \LTSMIN uses a 
 \emph{dependency matrix}, which has to be computed statically.

\begin{definition}[\PINS Matrix: \cite{blom2009:bridging}, Def. 4]
A \emph{dependency matrix} $D_{K \times N} = \mathit{DM}(P)$ for 
system $P$ is a matrix with $K$ rows
and $N$ columns containing $\set{0, 1}$ such that if $D_{k,i} = 0$ then 
group $k$ is independent of element $i$.\\
For any transition group $1 \leq k \leq K$, we define $\pi_k$ as the projection
$\pi_k \oftype S \to \Pi_{\set{1 \leq i \leq N \mid D_{k,i} = 1}} S_i$.
\end{definition}

\emph{Independence} here means that for given transition group $k$ the transitions 
do not depend on part $i$ of the state vector (\emph{read independence})
and the transitions do not change part $i$ of the successor state vector (\emph{write independence})
or that part $i$ is \emph{irrelevant} in both the current state and all successor states.
\emph{Irrelevant} here means that changing the value of that part would still
result in a bisimilar state space. For a more precise definition, see \cite[Def. 9]{vdpol2009:state}. 
This definition of independence is slightly more liberal than the one in \cite{blom2009:bridging} in that
we added this notion of relevance.

\subsubsection{Transition caching}

One way of exploiting the dependency information in the matrix is by using transition caching.
%\begin{algorithm}[htp]
%\caption{\Generate($s$). Generate the set of reachable states and the transitions between the states
%for an initial state $s_0$, given a next-state function $\nextst$.}
%\label{algorithm:reachability}
%\begin{algorithmic}[1]
%{\footnotesize
%  \STATE $S \becomes W \becomes \set{s_0}$
%  \STATE $T \becomes \emptyset$
%  \WHILE{$W \neq \emptyset$}
%    \STATE Choose $s \in W$
%    \STATE $S' \becomes \nextst(s) \setminus S$
%    \STATE $T \becomes T \cup \set{(s, s') \mid s' \in S'}$
%    \STATE $W \becomes (W \cup S') \setminus \set{s}$
%    \STATE $S \becomes S \cup S'$
%  \ENDWHILE
%  \RETURN $\tuple{S, T}$
%}
%\end{algorithmic}
%\end{algorithm}
\begin{algorithm}[tb]
\caption{\nextcache($s$, $k$) computes successors of $s$ for group $k$ using a cache.}
\label{algorithm:nextcache}
\centering
\begin{minipage}[t]{.32\linewidth}
\nextcache($s$, $k$)

\begin{algorithmic}[1]
{\footnotesize
  \STATE $\updatecache(s, k)$
  \STATE $S \becomes \emptyset$
  \FORALL{$t \in \cache_k[\pi_k(s)]$}
    \STATE $t' \becomes \nextapply(s, t, k)$
    \STATE Add $t'$ to $S$
  \ENDFOR
  \RETURN {$S$};
}
\end{algorithmic}
\end{minipage}
\hspace{.15cm}
\begin{minipage}[t]{.3\linewidth}
\updatecache($s$, $k$)

\begin{algorithmic}[1]
{\footnotesize
  \IF{$\pi_k(s) \notin \dom(\cache_k)$}
    \STATE $S \becomes \emptyset$
    \STATE $S' \becomes \groupnext(s,k)$
    \FORALL{$s' \in S'$}
      \STATE Add $\pi_k(s')$ to $S$
    \ENDFOR
    \STATE $\cache_k[\pi_k(s)] \becomes S$
  \ENDIF
}
\end{algorithmic}
\end{minipage}
\hspace{.15cm}
\begin{minipage}[t]{.25\linewidth}
\nextapply($s$, $t$, $k$)

\begin{algorithmic}[1]
{\footnotesize
  \STATE $j \becomes 1$
  \FOR{$1 \leq i \leq N$}
    \IF{$D_{k,i} = 0$}
      \STATE $s'[i] \becomes s[i]$
    \ELSE
      \STATE $s'[i] \becomes t[j]$
      \STATE $j \becomes j + 1$
    \ENDIF
  \ENDFOR
  \RETURN {$s'$};
}
\end{algorithmic}
\end{minipage}
\end{algorithm}%
Only the dependent parts of the transition are stored in a cache
 $\cache_k$ for every group $k$ by using the projection function $\pi_k$,
as described in \cite{blom2009:bridging} and shown in Alg.~\ref{algorithm:nextcache}.
This way time is saved, because
caching of transitions avoids calling $\groupnext$ at every step.
The density of the matrix has great influence on the performance of caching and of the 
symbolic tools. 

\subsection{PBES Language Module}

In this section we describe states, transition groups and the dependency
matrix for PPGs.
We assume to have a rewriter $\Simplify$ that is powerful enough to evaluate any closed data expression to $\ttrue$ or $\tfalse$ or to a disjunction or conjunction of predicate variables with closed data expressions as parameters. We use the same rewriter by Van Weerdenburg \cite{weerdenburg2009:efficient} as used in \cite{ploeger2011:verification}.

\subsubsection{States and transition groups}

For PPGs, the state vector is partitioned as follows:
$\tuple{\propvar{X}, x_1, x_2, \dotsc, x_M }$,
where $\propvar{X}$ is a propositional variable, and for $i \in \set{1 \ldots M}$ each $x_i$ is the value of parameter $i$. $M$ is the total number of parameter signatures in the system (consisting of name and type).

We assume the existence of a function $\textit{priority} \oftype S \to \Int$ that assigns a priority to each state
(based on the block of the corresponding equation) and a function 
$\textit{player} \oftype S \to \set{\Eloise, \Abelard}$ that assigns a player to each state (\Eloise if the corresponding
expression is a disjunction, \Abelard if it is a conjunction).
In particular, the $\ttrue$ state has priority $0$ and is owned by player $\Abelard$ and the $\tfalse$ state has priority $1$ and belongs to player $\Eloise$.

The equations in the PPG specify the transitions between states. 
The right hand side of the equation is split into conjuncts or disjuncts if possible,
which form the \emph{transition groups}, which are numbered subsequently.
We use a mapping $\var \oftype \Int \to \mathcal{X}$ from group number to variable and 
a mapping $\expression \oftype \Int \to \PF$ from group number to corresponding conjunct or disjunct.
In the following we assume the index sets $I$ and $J$ to be disjoint.

For a sequence of equations of the form
\[ \sigma \propvar{X}(\vec{d} \oftype D) = \Land_{i \in I} f_i \land
          \Land_{j \in J} \forall {\vec{v} \in D_j} \suchthat \big( g_j(\vec{d},\vec{v}) \impl \propvar{X}_j(e_j(\vec{d},\vec{v})) \big), \]
for each $i \in I$ there is a group $k$ with $\expression(k) = f_i$
and for each $j \in J$ there is a group $k$ with 
\begin{align*}
\expression(k) &= \forall {\vec{v} \in D_j} \suchthat \big( g_j(\vec{d},\vec{v}) \impl \propvar{X}_j(e_j(\vec{d},\vec{v})),
\end{align*}
and $\var(k) = \propvar{X}$. Symmetrically for disjunctive equations.

\begin{example}\label{example:buffer2}
We will explain these concepts using a specification of two sequential buffers (\texttt{buffer.2}):
\begin{quote}
\begin{tabular}{l@{}l@{\hspace{3pt}}l}
%$\sort\ $ & $D = \struct\ d1 \mid d2;$ \\
%$\act\ $  & $\action{r1}, \action{s4} \oftype D;$ \\
%$\act\ $  &	$\action{r}, \action{w}, \action{c} \oftype \Pos \# D;$ \\
%$\mcrlmap\ $  & $N \oftype \Pos;$ \\
%$\eqn\ $  & $N = 2;$ \\
$\proc\ $ & $\procname{In}(i: \Pos, q: \container{List}(D)) = $
        & $\displaystyle\sum_{d \oftype D}\ (\#q < 2) \guards \action{r_1}(d) \suchthat \procname{In}(i, q \append d)$ \\
      & & $\quad + \ (q \neq []) \guards \action{w}(i+1, \head(q)) \suchthat \procname{In}(i, \tail(q));$ \\

$\proc\ $ & $\procname{Out}(i: \Pos, q: \container{List}(D)) = $
        & $\displaystyle\sum_{d \oftype D}\ (\#q < 2) \guards \action{r}(i, d) \suchthat \procname{Out}(i, q \append d)$ \\
      & & $\quad + \ (q \neq []) \guards \action{s_4}(\head(q)) \suchthat \procname{Out}(i, \tail(q));$ \\

$\init\ $  
  & \multicolumn{2}{@{}l@{}}{$\hide(\{\action{c}\}, \allow(\{\action{r_1},\action{c},\action{s_4}\}, \comm(\{\action{w} \mid \action{r} \to \action{c}\}, \; \procname{In}(1,[]) \parallel \procname{Out}(2,[]) \; )));$}\\
\end{tabular}
\end{quote}\medskip
The specification of the initial state the system is specified as composed of an $\procname{In}$ and an $\procname{Out}$ component, composed with the \emph{parallel composition} ($\parallel$) operator. \emph{Synchronisation} of $\action{r}$ and $\action{w}$ actions of the two processes proceeds in two steps. The simultaneous occurence of actions $\action{r}$ and $\action{w}$ (the multi-action $\action{w} \mid \action{r}$) is renamed to $\action{c}$ ($\comm$) and 
separate occurances of $\action{r}$ and $\action{w}$ are ruled out by the \emph{restriction} operator ($\allow$).
The internal action $\action{c}$ is \emph{hidden} ($\hide$).
This specification is translated to a single process by \emph{linearising} it to Linear Process Specification (LPS) format. The result is the following specification:
\begin{quote}
\begin{tabular}{l@{}l@{\hspace{3pt}}l}
%$\sort\ $ & $D = \struct\ d1 \mid d2;$ \\
%$\act\ $  & $\action{r1}, \action{s4} \oftype D;$ \\
%$\act\ $  &	$\action{r}, \action{w}, \action{c} \oftype \Pos \# D;$ \\
%$\mcrlmap\ $  & $N \oftype \Pos;$ \\
%$\eqn\ $  & $N = 2;$ \\
$\proc\ $ & $\procname{P}(q_{in},q_{out} \oftype \container{List}(D)) = $
        & $\displaystyle\sum_{d \oftype D}\ (\#q_{in} < 2) \guards \action{r_1}(d) \suchthat \procname{P}(q_{in} \append d, q_{out})$ \\
      & & $\quad + \ (q_{out} \neq []) \guards \action{s_4}(\head(q_{out})) \suchthat \procname{P}(q_{in}, \tail(q_{out}));$ \\
      & & $\quad + \ (q_{in} \neq [] \land \#q_{out} < 2) \guards \ttau \suchthat \procname{P}(\tail(q_{in}), q_{out} \append \head(q_{in}))$ \\
      & & $\quad + \ \tdelta;$ \\
$\init\ $  
  & $\procname{P}([],[]);$
\end{tabular}
\end{quote}\medskip
The result of hiding the $\action{c}$ action is the internal \ttau transition in the third summand.
Actions that are not in the set $\set{\action{r_1},\action{c},\action{s_4}}$ are replaced by a \tdelta as a result of the restriction operator.\\
For this process specification, we want to verify the property that if a message is read through $\action{r_1}$, it will
 eventually be sent through $\action{s_4}$:
%\[ \always{\ttrue^\ast}(\forall d \oftype D \suchthat (\always{\action{r_1}(d)}(\mu \propvar{X} \suchthat (\possibly{\ttrue}\ttrue \land \always{\neg\action{s_4}(d)}\propvar{X})))) \]
\[ \nu \propvar{Y} \suchthat
(\forall d \oftype D \suchthat (\always{\action{r_1}(d)}(\mu \propvar{X} \suchthat (\possibly{\ttrue}\ttrue \land \always{\neg\action{s_4}(d)}\propvar{X})))) 
\land \always{\ttrue}\propvar{Y} \]
Satisfaction of this formula by the LPS translates to the following PBES:%
\begin{align}
\notag
\pbes\  & \nu \propvar{Y}(q_{in},q_{out} \oftype \container{List}(D)) = \\
        & \qquad\quad\quad (\forall_{d \oftype D} \suchthat (\#q_{in} < 2) \impl \propvar{X}(q_{in} \append d, q_{out}, d)) 
          \label{ex:pbes-groups:first} \\
        & \qquad\quad \land (\forall_{d_0 \oftype D} \suchthat (\#q_{in} < 2) \impl \propvar{Y}(q_{in} \append d_0, q_{out})) \label{ex:pbes-groups:yfirst} \\
\label{group:Y3} & \qquad\quad \land ((q_{out} \neq []) \impl \propvar{Y}(q_{in}, \tail(q_{out})))\\
        & \qquad\quad \land ((q_{in} \neq [] \land \#q_{out} < 2) \impl \propvar{Y}(\tail(q_{in}), q_{out} \append \head(q_{in}))); \label{ex:pbes-groups:ylast} \\
\notag
        & \mu \propvar{X}(q_{in},q_{out} \oftype \container{List}(D), d \oftype D) = \\
        & \qquad\quad\quad (\#q_{in} < 2) \lor (q_{out} \neq []) \lor (q_{in} \neq [] \land \#q_{out} < 2) 
        \label{ex:pbes-groups:truetrue} \\
        & \qquad\quad \land (\forall_{d_0 \oftype D} \suchthat (\#q_{in} < 2) \impl \propvar{X}(q_{in} \append d_0, q_{out}, d)) \label{ex:pbes-groups:xfirst}\\
        & \qquad\quad \land ( (\head(q_{out}) \neq d) \land (q_{out} \neq []) \impl \propvar{X}(q_{in}, \tail(q_{out}), d)) \\
        & \qquad\quad \land ( (q_{in} \neq [] \land \#q_{out} < 2) \impl \propvar{X}(\tail(q_{in}), q_{out} \append \head(q_{in}), d) ); \label{ex:pbes-groups:last}\\
\notag
\init\ & \propvar{Y}([], []);
\end{align}
For this equation system, the structure of the state vector is $\tuple{\propvar{X}, q_{in}, q_{out}, d}$.
The initial state would be encoded as $\tuple{\propvar{Y}, [], [], 0}$. Since the initial
state has no parameter $d$, a default value is chosen.
The numbers \ref{ex:pbes-groups:first}--\ref{ex:pbes-groups:last} behind the equation parts denote the different transition groups, i.e.,
each conjunct of a conjunctive expression forms a group.
For instance, for group \ref{group:Y3} the associated expression is $\expression(\ref{group:Y3}) = ((q_{out} \neq []) \impl \propvar{Y}(q_{in}, \tail(q_{out})))$ and it is associated with variable $\var(\ref{group:Y3}) = \propvar{Y}$.
Group \ref{ex:pbes-groups:first} encodes the $\always{\action{r_1}(d)}\varphi$ part of the formula (where $\varphi$ is the $\mu\propvar{X}$ part of the formula), groups \ref{ex:pbes-groups:yfirst}--\ref{ex:pbes-groups:ylast} encode the $\always{\ttrue}\propvar{Y}$ part,
group \ref{ex:pbes-groups:truetrue} encodes that a transition is enabled ($\possibly{\ttrue}\ttrue$), and groups \ref{ex:pbes-groups:xfirst}--\ref{ex:pbes-groups:last} encode the cases that not an $\action{r_4}(d)$ transition is taken.

\end{example}
\medskip

For an equation $\sigma \propvar{X}(\vec{d} \oftype D) = \varphi$, let $\params(\propvar{X})$ be the list of parameters $\vec{d}$
and $\params(\propvar{X})_i$ the $i$-th element of that list.
The next state function $\groupnext$ is defined as follows.
For every $k$ with $\var(k) = \propvar{X}$,
\[ \groupnext(\propvar{X}(\vec{e}), k) \eqdef \begin{cases}
\set{ \Simplify(f[\params(\propvar{X}) \becomes \vec{e}]) } \\
  \qquad \text{ if $f = \expression(k)$ is a simple formula; } \\
\set{ \propvar{X}'(h(\vec{e},\vec{v})) \mid \vec{v} \in D \land g(\vec{e},\vec{v}) } \\
  \qquad \text{ if } \expression(k) \text{ is of the form }
 \mathsf{Q} {\vec{v} \in D} \suchthat \big( g(\vec{e},\vec{v}) \oplus \propvar{X}'(h(\vec{e},\vec{v}))
\end{cases} \]
Note that if $f$ is a simple expression, $\Simplify(f[\params(\propvar{X}) \becomes \vec{e}])$ will result in either \ttrue or \tfalse.
In the case that $f$ is not simple, all concrete variable instantiations are enumerated for every quantifier variable $\vec{v}$ for which the guard $g$ is satisfied.

\begin{example}
For the example above, 
$\groupnext(\propvar{Y}([], []), \ref{group:Y3})$ yields the empty set because $q_{out} = []$.
$\groupnext(\propvar{Y}([], []), \ref{ex:pbes-groups:yfirst})$ results in $\set{\propvar{Y}([d_1], []), \propvar{Y}([d_2], [])}$.
\end{example}

\subsubsection{Dependency matrix}

Let $\occ(\varphi)$ be the set of propositional variable occurring in a term $\varphi$,
let $\free(d)$ be the set of \emph{free data variables} occurring in a data term $d$, and
$\used(\varphi)$ the 
set of free data variables occurring in an expression $\varphi$ such that
the variables are not merely passed on to the next state.
E.g., with $\propvar{X}(a, b) = \xi$, for the expression $\varphi = a \land \propvar{X}(c, b)$,
$\used(\varphi) = \set{a, c}$. Parameter $b$ is not in the set because it does not influence the computation,\
but is only passed on to the next state.
For a formula $\varphi$, the function $\changed(\varphi)$
computes the variable parameters changed in the formula:
\[
 \changed(\propvar{X}(e_1, \ldots, e_m))   \eqdef\eqsep \set{d_i \mid i \in \set{1 \ldots m} \land d_i = \params(\propvar{X})_i \land e_i \neq d_i}
\]
The function $\booleanResult(\varphi)$
determines if $\varphi$ contains a branch that directly results in a \ttrue or \tfalse (not a variable).
This is needed because the boolean constants are encoded as a vector with variable names ``true'' and ``false'', hence a transition to one of them changes the first part of the state vector.
For group $k$ and part $i$, we define read dependence $d_{R}$ and write dependence $d_{W}$:
\begin{align*}
d_{R}(k, i) &\eqdef \begin{cases}
\ttrue & \text{ if } i=1;\\
p_i \in ( \params(\var(k)) \cap \used(\expression(k)) ) & \text{ otherwise. }
\end{cases}\\
d_{W}(k, i) &\eqdef \begin{cases}
\left( \occ(\expression(k)) \setminus \set{\var(k)} \neq \emptyset \right) 
\; \lor \; \booleanResult(\expression(k)) 
& \text{ if } i=1;\\
p_i \in \changed(\expression(k))
& \text{ otherwise. }
\end{cases}
\end{align*}
$d_{R}(k, 1)$ is true for every group $k$, since the variable has to be read to determine
if a transition group is applicable.

\begin{definition}[PPG Dependency matrix]\label{def:dependencymatrix}
For a PPG $P$ the dependency matrix $\mathit{DM}(P)$ is a $K \times M$ matrix
defined for $1 \leq k \leq K$ and $1 \leq i \leq M$ as:
\begin{align*}
\mathit{DM}(P)_{k,i} &= 
\begin{cases}
+ & \text{ if } d_{R}(k, i) \land d_{W}(k, i); \\
r & \text{ if } d_{R}(k, i) \land \neg d_{W}(k, i); \\
w & \text{ if } \neg d_{R}(k, i) \land d_{W}(k, i); \\
- & \text{ otherwise. }
\end{cases}
\end{align*}
\end{definition}

\begin{example}
For the PBES in Example~\ref{example:buffer2}, the dependency matrix looks like this:\\
\begin{tabular*}{\textwidth}{@{}c@{\extracolsep{\fill}}m{4in}@{}}
%\scalebox{1}{
\begin{tabular}{l|cccc}
$k$ & $\propvar{X}$ & $q_{in}$ & $q_{out}$ & $d$\\
\hline
$1$&$+$&$+$&$-$&$w$\\
$2$&$+$&$+$&$-$&$-$\\
$3$&$+$&$-$&$+$&$-$\\
$4$&$+$&$+$&$+$&$-$\\
$5$&$+$&$r$&$r$&$-$\\
$6$&$+$&$+$&$-$&$-$\\
$7$&$+$&$-$&$+$&$r$\\
$8$&$+$&$+$&$+$&$-$\\
\end{tabular}
%}
& \smallskip
The first row lists the state vector parts. The left column lists the group numbers. A `$+$' denotes both read and write dependency, `$w$' denotes write dependency, `$r$' read dependency, and `$-$' no dependency between the group and the state vector part.
For group \ref{ex:pbes-groups:first} we can see that the variable is changed from $\propvar{Y}$ to $\propvar{X}$, which results in a `$+$' in the $\propvar{X}$ column. The $q_{in}$ parameter is both read and changed ($d$ is added to it). The $q_{out}$ parameter is not touched, which results in a `$-$'. The parameter $d$ is not in 
$\params({\propvar{Y}})$ and therefore there is no read dependence. However, the value of $d$ is set for the next state, resulting in a `$w$' in the last column.
\end{tabular*}
\end{example}

\section{Performance Evaluation}\label{section:experiments}

In this section we report the performance of our tools compared to existing tools in the
\MCRLTWO toolset. 

\subsection{Experiment setup}

As input we used PBESs that are derived from the following \MCRLTWO models:
$n$ sequential buffers (\texttt{buffer-*}),
the Sliding Window Protocol (SWP),
the IEEE 1394 protocol,
a Sokoban puzzle, and
state machines that are part of the control system for an experiment at CERN (\texttt{wheel\_sector}), described in \cite{hwong2011:analysing}.
The models are combined with \MUCALC properties that check
absence of deadlock (\texttt{nodeadlock}, see Example~\ref{example:buffer}), 
if $x$ is read, then eventually $x$ will be written (\texttt{evt\_send}, see Example~\ref{example:buffer2}), or
that from the initial state there is a path on which a $\action{push}$ action is possible (\texttt{always\_push}: 
$\possibly{\ttrue^\ast}\possibly{\action{push}}\ttrue$ --
only applicable to the Sokoban puzzle).

As preprocessing steps,
we applied \texttt{pbesparelm} and \texttt{pbesrewr -psimplify} to every equation system, which are rewriters that apply some obvious simplifications to the equation systems.
In the reported cases no transformation to PPG was needed, as the systems were already in
the required form.
%\texttt{buffer} with property~\ref{property:read-then-write} required (manually) replacing expressions of the form $\ttrue \land \varphi$ by $\ttrue$ and rewriting expressions
%of the form $\forall d_1 \oftype D \suchthat \forall d_2 \oftype D \suchthat (d_1 = d_2) \implies \varphi$ to
%$\forall d_1 \oftype D \suchthat \varphi[d_2 \becomes d_1]$ (if sort $D$ is not empty) to be in PPG form.

The tools that we compared are:
\begin{center}
\scalebox{0.9}{
\begin{tabular}{l c c c c c l}
\toprule
Tool & Toolset & \begin{sideways}Groups\end{sideways} & \begin{sideways}Caching\end{sideways} & \begin{sideways}Distributed\end{sideways} &\begin{sideways}Symbolic\end{sideways} & Command \\
\toprule
\texttt{pbes2bes}         & \MCRLTWO & -- & -- & -- & --      & \texttt{pbes2bes -rjittyc} \\
\midrule
\texttt{pbespgsolve}      & \MCRLTWO & -- & -- & -- & --      & \texttt{pbespgsolve -rjittyc -g} \\
\midrule
\texttt{pbes2lts -black}  & \LTSMIN & no & no & no & no       & \texttt{pbes2lts-grey {-}-black {-}-always-split} \\
\midrule
\texttt{pbes2lts -grey}   & \LTSMIN & yes & no & no & no      & \texttt{pbes2lts-grey {-}-grey {-}-always-split} \\
\midrule
\texttt{pbes2lts -cache}  & \LTSMIN & yes & yes & no & no     & \texttt{pbes2lts-grey -rgs -c {-}-always-split} \\
\midrule
\texttt{pbes2lts-mpi-*}   & \LTSMIN & yes & yes & yes & no    & \texttt{pbes2lts-mpi -rgs -c {-}-always-split} \\
\midrule
\texttt{pbes-reach}       & \LTSMIN & yes & no & no & yes     & \texttt{pbes-reach {-}-order=chain-prev} \\
                          &         & & & &                   & \quad \texttt{{-}-saturation=sat-like} \\
                          &         & & & &                   & \quad \texttt{{-}-save-levels -rgs} \\
                          &         & & & &                   & \quad \texttt{{-}-always-split} \\  
\bottomrule
\end{tabular}
}
\end{center}
It is indicated whether transition groups, caching, distributed generation or symbolic generation are available.
\texttt{pbes2bes} and \texttt{pbespgsolve} from the \MCRLTWO toolset are similar in functionality, but different in implementation.
For \texttt{pbespgsolve} the \texttt{-g} option means only generating the parity game without solving.
For the \LTSMIN tools \texttt{pbes2lts-*} and \texttt{pbes-reach} the option \texttt{-rgs} enables regrouping, \texttt{-c} enables caching, and \texttt{{-}-black}
disables the use of transition groups. \texttt{pbes-reach} uses the \texttt{sat-like} saturation strategy.

The experiments were performed on a cluster of 10 machines with each two quad-core Intel Xeon E5520 CPUs @ 2.27 GHz (with 2 hyperthreads per core) and 24GB
memory. Every tool was given a 20 GB memory limit and a 10 ks time limit. Elapsed time and memory usage have been measured by the tool \texttt{memtime}.
The experiments were executed using Linux 2.6.34, \MCRLTWO svn rev. 10785 and for \LTSMIN the git rev. after commit 4d11bc20 in the experimental `next' branch. The tools were built using GCC 4.4.1. Open MPI 1.4.3 was used for the distributed tool.

\begin{table}
\caption{Time performance in seconds. `T' indicates a timeout, `M' out of memory.}
\label{table:results-time}
\begin{center}
\scalebox{0.8}{
\begin{tabular}{lrrrrrrrrrr}
\toprule
Equation system & \# States & \begin{sideways}\texttt{pbes2bes                 }\end{sideways} & \begin{sideways}\texttt{pbespgsolve              }\end{sideways} & \begin{sideways}\texttt{pbes2lts -black           }\end{sideways} & \begin{sideways}\texttt{pbes2lts -grey            }\end{sideways} & \begin{sideways}\texttt{pbes2lts -cache    }\end{sideways} & \begin{sideways}\texttt{pbes2lts-mpi-1           }\end{sideways} & \begin{sideways}\texttt{pbes2lts-mpi-4           }\end{sideways} & \begin{sideways}\texttt{pbes2lts-mpi-8           }\end{sideways} & \begin{sideways}\texttt{pbes-reach        }\end{sideways} \\
\toprule
\texttt{swp.nodeadlock           }
 &        1,862 &       5 &       5 &       5 &       5 &       5 &       5 &       5 &       5 &       5 \\
\midrule
\texttt{swp.evt\_send            }
 &       33,554 &       7 &       7 &       8 &      11 &       5 &       5 &       5 &       8 &       5 \\
\midrule
\texttt{1394.nodeadlock          }
 &      173,101 &     199 &     202 &     231 &   1,387 &     120 &     125 &      56 &      73 &     114 \\
\midrule
\texttt{sokoban.372.always\_push }
 &      834,397 &      69 &      78 &     258 &       T &     403 &     419 &     182 &      62 &      31 \\
\midrule
\texttt{buffer.7.nodeadlock      }
 &      823,545 &      32 &      33 &      48 &      76 &      13 &      16 &       9 &       7 &       9 \\
\midrule
\texttt{buffer.7.evt\_send       }
 &    2,466,257 &     111 &     107 &     157 &     266 &      22 &      27 &      13 &      11 &       9 \\
\midrule
\texttt{buffer.8.nodeadlock      }
 &    5,764,803 &     235 &     237 &     357 &     594 &      82 &      93 &      31 &      20 &      37 \\
\midrule
\texttt{buffer.8.evt\_send       }
 &   17,281,283 &     820 &     859 &   1,256 &   2,171 &     158 &     191 &      71 &      67 &      42 \\
\midrule
\texttt{buffer.9.nodeadlock      }
 &   40,353,607 &   1,059 &       M &   2,937 &   4,905 &     571 &     686 &     241 &     197 &     274 \\
\midrule
\texttt{buffer.9.evt\_send       }
 &  121,021,455 &       M &       M &       T &       T &   1,172 &   1,448 &     520 &     306 &     282 \\
\midrule
\texttt{wheel\_sector.nodeadlock }
 &    4,897,760 &       T &       T &       T &       T &   2,337 &   2,368 &     828 &     939 &   1,904 \\
\bottomrule
\end{tabular}
}
\end{center}
\end{table}

\begin{table}
\caption{Memory usage in MB. `T' indicates a timeout, `M' out of memory.}
\label{table:results-mem}
\begin{center}
\scalebox{0.8}{
\begin{tabular}{lrrrrrrrrrr}
\toprule
Equation system & \# States & \begin{sideways}\texttt{pbes2bes                 }\end{sideways} & \begin{sideways}\texttt{pbespgsolve              }\end{sideways} & \begin{sideways}\texttt{pbes2lts -black           }\end{sideways} & \begin{sideways}\texttt{pbes2lts -grey            }\end{sideways} & \begin{sideways}\texttt{pbes2lts -cache    }\end{sideways} & \begin{sideways}\texttt{pbes2lts-mpi-1           }\end{sideways} & \begin{sideways}\texttt{pbes2lts-mpi-4           }\end{sideways} & \begin{sideways}\texttt{pbes2lts-mpi-8           }\end{sideways} & \begin{sideways}\texttt{pbes-reach        }\end{sideways} \\
\toprule
\texttt{swp.nodeadlock           }
 &        1,862 &        12 &        11 &        17 &        17 &        16 &        13 &        15 &        14 &        16 \\
\midrule
\texttt{swp.evt\_send            }
 &       33,554 &        58 &        29 &        20 &        20 &        18 &        15 &        15 &        16 &        47 \\
\midrule
\texttt{1394.nodeadlock          }
 &      173,101 &       227 &       168 &        31 &        30 &        89 &        86 &        60 &        50 &        57 \\
\midrule
\texttt{sokoban.372.always\_push }
 &      834,397 &     1,187 &       768 &        34 &       T &       220 &       217 &        69 &        45 &        47 \\
\midrule
\texttt{buffer.7.nodeadlock      }
 &      823,545 &       965 &       354 &        32 &        32 &        91 &        89 &        36 &        27 &        49 \\
\midrule
\texttt{buffer.7.evt\_send       }
 &    2,466,257 &     3,340 &     1,215 &        63 &        64 &       181 &       179 &        67 &        43 &        49 \\
\midrule
\texttt{buffer.8.nodeadlock      }
 &    5,764,803 &     7,179 &     2,579 &       117 &       117 &       528 &       525 &       145 &        81 &        49 \\
\midrule
\texttt{buffer.8.evt\_send       }
 &   17,281,283 &    18,136 &     9,056 &       345 &       345 &     1,155 &     1,152 &       377 &       204 &        49 \\
\midrule
\texttt{buffer.9.nodeadlock      }
 &   40,353,607 &    18,451 &       M &       737 &       737 &     4,129 &     4,127 &     1,048 &       538 &        49 \\
\midrule
\texttt{buffer.9.evt\_send       }
 &  121,021,455 &       M &       M &       T &       T &     9,209 &     9,206 &     3,003 &     1,487 &        49 \\
\midrule
\texttt{wheel\_sector.nodeadlock }
 &    4,897,760 &       T &       T &       T &       T &     1,288 &     1,285 &       389 &       238 &        90 \\
\bottomrule
\end{tabular}
}
\end{center}
\end{table}

\subsection{Results}
Results are in Tables~\ref{table:results-time} (time performance in seconds) and \ref{table:results-mem} (memory usage in MB). For the MPI tool, the values are the maximum for the workers. The `T' indicates a timeout, the `M' indicates an Out of Memory error.
We can make the following observations.

From the results we see that \texttt{pbes2bes} and \texttt{pbespgsolve} from the \MCRLTWO toolset perform better than
\texttt{pbes2lts -black}, the \LTSMIN based tool without any optimisation.
The memory performance of the \LTSMIN tool however is much better, even over 25 times better in the case of
\texttt{buffer.8.evt\_send}.

Looking at \texttt{pbes2lts -grey} we observe that only splitting into transition groups without any optimisations 
has a negative impact on the performance, especially in the case of \texttt{1394.nodeadlock}.

The \LTSMIN tools have a relatively bad performance for the Sokoban puzzle, because of the structure of
\texttt{always\_push}: either ``we can do a push now'' or ``we move and take a recursive step''.
If this formula is evaluated as a whole on a state where we can do a push, the first part will immediately evaluate to \ttrue and the formula as well, without taking the recursive step.
When the formula is split into transition groups, then both parts may be evaluated independently. Although the second part is not needed, such on-the-fly solving optimisations are not available in the PBES language module yet when transition groups are enabled.
This causes \LTSMIN to generate a state space of 10,992,856 states (instead of 834,397), but still the symbolic tool of \LTSMIN, \texttt{pbes-reach}, is the fastest.

Transition caching pays off for many systems. Compared to the \MCRLTWO tools, the speedup is between 1.8 and 5.1 for the
sequential buffers and for \texttt{wheel\_sector} the instantiation is completed within the timebound.
The distributed tool does not scale well. The speedup with 8 workers compared to 1 worker is 6.8 for the Sokoban puzzle, but does not exceed 4.7 for the sequential buffers, and is only 2.5 for the \texttt{wheel\_sector} case.
In the \texttt{wheel\_sector} and \texttt{1394} cases
the execution time for 8 workers is even worse than with 4 workers, indicating that
there is a limit to the number of workers that result in a further speedup. 

The symbolic tool performs best of all sequential tools in all cases.
The tool is up to 19.5 times faster than the fastest tool from the \MCRLTWO toolset (in the \texttt{buffer.8.evt\_send} case). And for some cases \LTSMIN could finish within memory and time bounds, whereas the \MCRLTWO tools could not. Memory usage of \texttt{pbes-reach} is slightly worse in the smallest cases, but up to more than 180 times better than the \MCRLTWO tools for the other cases.

%\section{Discussion}

%\cite{mateescu2008:model}.

\section{Conclusions}\label{section:conclusions}

We have defined PPG as normal form for PBESs and a transformation to PPG, 
making the instantiation to parity games more straightforward.
We implemented a PBES language module for \LTSMIN. As a result, 
the high-performance capabilities for state space generation become
available for parity game generation. We demonstrated this for distributed
state space generation and for symbolic state space generation.

Experimental comparison to existing tools shows good results. 
The \LTSMIN tools reduce memory usage enormously.
Transition caching, distributed computation and the symbolic tool
speed up the instantiation in all reported cases. 
However, the distributed tool does not scale well. 
For all reported cases, the symbolic \LTSMIN tool performed the best, with up to 19 times speedup 
and up to more than 180 times lower memory usage compared to the \MCRLTWO tools.

We intend to extend the tool with optimisations, such as on-the-fly minimisation and solving, i.e., while generating the parity game (possibly also distributed).
Furthermore, the symbolic tool generates a BDD representation of the parity game,
which asks for solvers that can deal with such symbolic parity games similar to the tool by \cite{bakera2009:solving}.

\vspace*{-\medskipamount}
\paragraph{Acknowledgments.}
We are grateful to Tim Willemse, Jeroen Keiren and Wieger Wesselink for their support on the \MCRLTWO toolset.

%\nocite{bergstra2001:handbook}
\bibliographystyle{plain}
\bibliography{bibliography/kant-phdproject}

\end{document}